\documentclass[aps,preprint,12pt,onecolumn]{revtex4}%
\usepackage{amsfonts}
\usepackage{amsmath}
\usepackage{amssymb}
\usepackage{graphicx}%
\providecommand{\U}[1]{\protect\rule{.1in}{.1in}}
\begin{document}
\preprint{cond-mat 01/2008}
\title[$CaFeAs$]{Character of magnetic instabilities in CaFe$_{2}$As$_{2}$}
\author{G. D. Samolyuk and V. P. Antropov}
\affiliation{Condensed Matter Physics, Ames Laboratory, IA 50011}
\pacs{PACS number}

\begin{abstract}
The density functional non-interacting susceptibility has been analyzed in
different phases of CaFe$_{2}$As$_{2}$ and compared with similar data for pure
d-metals. The conditions for the "no local moment" itinerant state with large
frustrations are found for the "collapsed" phase (corresponding to
superconducting phase). This itineracy determines the instability versus the
incommensurate magnetic order for the narrow region of wave vectors. For the
ambient pressure phase, the local moments on Fe atoms with much less
frustrated antiferromagnetic interactions are stabilized and a magnetic short
or long range order is developed.

\end{abstract}
\eid{identifier}
\date{\today}
\maketitle

The metallic state and magnetic properties of new Fe-As based
superconductors\cite{Kamihara,Caexp} raised the question about a nature of
magnetic interactions in these systems. Measured magnetic moments are small
($\sim$0.3-0.8 $\mu_{B}$) and indicated that these materials most likely
itinerant and are close to non magnetic state. Similar conclusions have been
reached in band structure studies\cite{Singh}. However specific details of
magnetic instabilities in these systems have not been analyzed, and so it is
unclear to what extent these systems are itinerant or localized. In this
report we analyze the different phases of CaFe$_{2}$As$_{2}$
superconductor\cite{Caexp} and estimate the criteria of magnetic instability
in real and reciprocal space. By comparing obtained results with similar
calculations for already known magnetic systems, we provide some additional
illustrations why a CaFe$_{2}$As$_{2}$ system can be classified as
antiferromagnetic (AFM) system at the borderline between itinerant and
localized behavior, with a degree of magnetic short or long range instability
determined primarily by Fe-As bonding.

The calculations have been done using local spin density approximation (LSDA)
with FLAPW, FPLMTO and TBLMTO-ASA methods. All electronic structure
calculations are in good agreement between each other and ASA results are very
close to the full-potential ones. The non-interacting susceptibility has been
calculated using an ASA-Green function\cite{EX1} approach%
\begin{equation}
\chi_{ij}=\frac{1}{\pi}Tr\operatorname{Im}%
%TCIMACRO{\dint }%
%BeginExpansion
{\displaystyle\int}
%EndExpansion
d\varepsilon G_{ij}\left(  \varepsilon\right)  G_{ji}\left(  \varepsilon
\right)  \label{a1}%
\end{equation}
where $G_{ij}\left(  \varepsilon\right)  $ is the Green function. Evidently
the sum rule for the density of states (DOS) is $N_{i}\left(  \varepsilon
\right)  =\sum\chi_{ij}$ and usual the Stoner criteria for ferromagnetism (FM)
is $IN\left(  \varepsilon_{F}\right)  =1,$with $I$ being a Stoner parameter.

The local moment criterion\cite{Ander} is written as
\begin{equation}
S_{0}=I\chi_{00}>1, \label{a2}%
\end{equation}
while criteria of the FM or AFM pair formation\cite{Moriya} (short range order
(SRO) instability) are%
\begin{equation}
S_{01}^{\pm}=I\left(  \chi_{00}\pm\chi_{01}\right)  >1, \label{a3}%
\end{equation}
where "+" is for FM and "-" is for AFM orderings. We also use generalized
stability parameter%
\begin{equation}
S_{N}=I\overset{N}{\underset{j=0}{%
%TCIMACRO{\dsum }%
%BeginExpansion
{\displaystyle\sum}
%EndExpansion
}}r_{j}\chi_{0j}, \label{a4}%
\end{equation}
with r$_{j}$="+" for FM orientation of moments $i$ and $j$ and r$_{j}$="-" for
AFM orientation.

We analyze the local and non-local susceptibility in real space for the
presumably itinerant systems and the natural question arises: how reliably can
the real space nearest neighbor (NN) coupling analyses predict the itinerant
magnetic state? To demonstrate the applicability of such an approach, we first
analyze several well known magnetic materials with a different degree of
itineracy (see Tab.I and Fig.1). Several typical magnetic scenarios can be
identified. BCC Fe represents the relatively localized magnet with criterion
(\ref{a2}) well satisfied (for the value of $I$ see Ref.\cite{Janak}). The
"itineracy" parameter $\alpha=\left(  S_{\infty}-S_{0}\right)  /S_{0}$ is just
$0.1,$ so BCC Fe is a local moment system with a well established long range
order (LRO) ($I\chi\left(  \mathbf{q=0}\right)  >1$) and small amount of
itineracy. The cases of Ni, Cr or fcc Mn are intermediate. The criterion
(\ref{a2}) is not satisfied while NN couplings support a developing
instability against LRO appearance. These systems can be classified as "no
local moment" or strongly itinerant ($\alpha>0.5$) systems with well
established LRO. The Mn case is special due to different results for BCC and
FCC structures. We show the results for no local moment FCC Mn ($\alpha
\simeq0.15$) while BCC Mn can form local moment. The FCC Pd represents a
system where local moment instability is highly unlikely ($S_{0}<0.4$) with a
large "itineracy" parameter $\alpha>0.5.$ NN susceptibilities, however are
also large and positive, and provide a corresponding increase of the DOS at
the Fermi level. Thus, the FCC Pd comes close to FM instability. The HCP Ti is
also non-magnetic and not very far from AFM unstability, with large negative
NN susceptibilities and all of the criteria (\ref{a2}-\ref{a3}) are not
fulfilled. All of these results are well supported by experiment, and provide
necessary justification of our real space analysis in other metallic systems.

Continuing with the case of CaFe$_{2}$As$_{2}$, Table II presents the local
and non-local susceptibilities for the two states of CaFe$_{2}$As$_{2}$ known
from the experiment as the superconducting (finite pressure) and normal
(ambient pressure) phases\cite{Caexp} with corresponding distances between Fe
and As atoms R$_{Fe\text{-}As}$.

Table II and Fig.2 show that CaFe$_{2}$As$_{2}$ represents a magnet at the
borderline between localized and itinerant behavior. The following picture
emerges from studies of non-magnetic susceptibility in these phases. At a
small R$_{Fe\text{-}As}$, (a superconducting phase with smaller volume) the
condition Eq.\ref{a2} is not fulfilled with $\chi_{00}$ being nearly twice
smaller than in BCC Fe (Table I). Thus the system can be classified as an
itinerant system with no local moment. However, the nonlocal susceptibility of
the two nearest neighbors (NN) is not small ($\alpha=\left(  S_{\infty}%
-S_{0}\right)  /S_{0}\sim0.19$; $\eta=\chi_{02}/\chi_{01}\simeq0.42$) and is
negative ($\chi_{01}<0,\chi_{02}<0$). So, the criterion for AFM pair formation
(\ref{a3}) along (1,0), (0,1) and (1,1) comes close to the instability
threshold and the corresponding SRO can be stabilized. Simultaneously, the
criterion in reciprocal space at certain points in $\mathbf{q}$-space also
appears to be close to instability (Fig.3). The criterion of FM pair formation
$S_{12}^{+}<1$ is not supported by NN coupling. FM fluctuations are thus
strongly suppressed while AFM interactions are frustrated. A magnetic moment
appears as a consequence of the sum of exchange fields from all sites,
supporting the itinerant nature of this ground state. For small R$_{Fe\text{-}%
As}$, the frustration parameter $\eta\simeq0.4$, for intermediate
R$_{Fe\text{-}As}$ $\eta\simeq0.2$ and at large R$_{Fe\text{-}As}$ it is even
smaller. This is related to the small value of $\chi_{01}$ for the normal
state. Such closeness to the zero (and to sign inversion) leads to large
anisotropy of the NN exchange parameters in magnetic state\cite{Kir, FESE}. As
R$_{Fe-As}$ increases, $I\chi\left(  \mathbf{q}\right)  $ approaches 1 for a
larger region of wave vectors (Fig.3), and with a further increase of
R$_{Fe\text{-}As}$, it is fulfilled even for $\mathbf{q}=0$ (FM LRO) (for
comparison see Ref.\cite{LIND, JAP} for calculations of $\chi\left(
\mathbf{q}\right)  $ and related functions in similar materials). At this
point the strong NN susceptibilities are not all negative and this state is no
longer reflecting small R$_{Fe-As}$ frustrations. The main difference between
the shape of $\chi\left(  \mathbf{q}\right)  $ in these phases is a
disappearance a maximum of $\chi\left(  \mathbf{q}\right)  $ at $\mathbf{q}$
corresponding to the stripe AFM structure and stabilization of non-collinear
state (Fig.3b). Our results are similar to those obtained in Ref.\cite{JAP}
where the susceptibility was calculated for undoped and doped cases of
LaFeAsO, while our results are for CaFe$_{2}$As$_{2}$ under different
pressure. Despite these differences, both calculations revealed very similar
trend: the strong stripe AFM instability for the normal phase and the
instability with respect to formation of non-collinear state for the system
where the superconductivity was observed experimentally. While the results are
similar the non-collinearity in our case is somewhat stronger. This
non-collinearity is directly related to the strong coupling between moments at
larger distances (beyond first two NN) which is very natural for the itinerant magnet.

One can parametrize the stability function $\chi\left(  \mathbf{q}\right)  $
using only Fe-Fe $\chi_{ij}$ from Tab.2:%
\begin{equation}
\chi\left(  \mathbf{q}\right)  =\chi_{01}\left(  \cos(q_{x}\right)
+\cos\left(  q_{y}\right)  )+2\chi_{02}\cos(q_{x})\cos\left(  q_{y}\right)
+\chi_{03}\left(  \cos(2q_{x}\right)  +\cos\left(  2q_{y}\right)  )+...
\label{a7}%
\end{equation}

While this parametrization reflects many important features of total
$\chi\left(  \mathbf{q}\right)  $ from Fig.3, it does not include, for
instance, Fe-As contributions.

In both phases, the shape of maximum of $\chi\left(  \mathbf{q}\right)  $ is
never sharp. The value R$_{Fe\text{-}As}$=2.328 \AA \ can be considered a
critical value (R$_{crit}$) for the local magnetic moment to be stable. The
critical value R$_{crit}$ of magnetic instability appears to be close to the
R$_{crit}$ values where the superconductivity has been observed\cite{Caexp}.
At intermediate R$_{Fe-As}$ there are regions where the local moment and LRO
criteria are not fulfilled while SRO criterion for AFM pairs $S_{12}^{-}$
$\geq1$. This peculiar phase of AFM NN pairs of Fe atoms exists in very small
interval of R$_{Fe-As}$. While the system is close to be classified as
itinerant, the formation of magnetism is somewhat unusual. The degree of
itinerancy is controlled by the competition of the strong tendency of
intraatomic exchange on the Fe atom to form localized moments (with the direct
exchange coupling between them), and Fe-As bonding destroying intraatomic
magnetic instability and adding superexchange to the pool of competing
interactions\cite{Kir}. The present estimations are based on non-magnetic
calculations, and the calculations for the actual magnetic state can reveal
somewhat renormalized behavior\cite{FESE}. Also, for simplicity we used the
value of Stoner parameters from Ref.\cite{Janak}, alternative estimations may
provide somewhat different numbers. These estimations do not include
zero-point motion effects or anharmonicity of spin fluctuations, which are
important in the strongly itinerant case. The studies of the dynamic nature of
local moments will be done in our following publication.

Our calculations revealed that $\chi_{02}$ is rather stable as a function of
R$_{Fe\text{-}As}$, while $\chi_{01}$ is very sensitive to that distance and
even changes its sign at larger R$_{Fe\text{-}As}$. This supports the view
that the crystallographic phases corresponding to normal and superconducting
states of CaFe$_{2}$As$_{2}$ have very different structures of pair
interactions, providing no justification for the applicability of the $J_{1}%
$-$J_{2}$ model. Thus, the "collapsed" tetragonal (superconducting) phase is
marginally itinerant "no local moment system", on the brink of an instability
against forming a short or long range non-collinear magnetic order and has all
prerequisites for strong frustrations between its first two nearest neighbors,
while the ambient pressure phase has a well defined static local magnetic
moment on Fe atoms with a stable short or long range magnetic order of a
stripe type.

V.A. would like to thank S.Bud'ko and K.Belashchenko for continuing inspiring
discussions. Work at the Ames Laboratory was supported by the Department of
Energy-Basic Energy Sciences, under Contract No. DE-AC02-07CH11358.

\begin{tabular}
[c]{|l|l|l|l|l|l|l|l|}\hline
& Ti & Cr & Mn & Fe & Co & Ni & Pd\\\hline
$N\left(  E_{F}\right)  $ & 13.5 & 10.5 & 21.5 & 43.9 & 31.5 & 56.1 &
32.1\\\hline
$\chi_{00}$ & 21.2 & 23.0 & 26.7 & 47.4 & 30.2 & 25.3 & 15.2\\\hline
$\chi_{01}$ & -0.37 & -1.3 & -0.54 & 1.5 & 0.74 & 1.8 & 0.94\\\hline
$\chi_{02}$ & -0.47 & -0.12 & 0.04 & -2.23 & -0.46 & 0.01 & 0.05\\\hline
$\chi_{03}$ & -0.31 & -0.05 & -0.11 & 0.27 & 0.18 & 0.32 & 0.14\\\hline
$\chi_{04}$ & 0.15 & -0.08 & 0.15 & -0.18 & 0.18 & 0.33 & 0.19\\\hline
S$_{0}$ & 0.53 & 0.64 & 0.80 & 1.61 & 1.09 & 0.92 & 0.36\\\hline
\end{tabular}

Table I. The density of states $N\left(  E_{F}\right)  $ at the Fermi level,
local and several non-local susceptibilities (in units of 1/Ry) in systems
with different character of magnetic coupling. The data have been obtained for
Ti and Co in HCP, Cr and Fe in BCC, Mn, Ni and Pd in FCC structures.

\begin{tabular}
[c]{|l|l|l|l|l|}\hline
& $n$ & $\overrightarrow{\tau}$ & N & S\\\hline
$\chi_{00}$ & 1 & $0,0,0$ & 34.4 & 27.1\\\hline
$\chi_{01}$ & 4 & $0.5,0,0$ & -0.09 & -0.99\\\hline
$\chi_{02}$ & 4 & $0.5,0.5,0$ & -0.53 & -0.42\\\hline
$\chi_{03}$ & 4 & $1,0,0$ & 0.35 & 0.07\\\hline
$\chi_{04}$ & 2 & $0,0,1$ & -0.09 & -0.06\\\hline
$\chi_{05}$ & 8 & $0.5,1,0$ & 0.12 & 0.16\\\hline
S$_{0}$ &  &  & 1.17 & 0.92\\\hline
\end{tabular}

Table II. The local and non-local Fe atom susceptibilities (in units of 1/Ry)
in CaFe$_{2}$As$_{2}$. 'S' - "collapsed" tetragonal\cite{Caexp}
(R$_{Fe\text{-}As}$=2.336 \AA ) and "N" -ambient pressure normal
(R$_{Fe\text{-}As}$=2.373 \AA ) phases. Column $n$ denotes the number of
equivalent nearest neighbors. $\overrightarrow{\tau}$ is the connecting vector
in units of the lattice parameter $a$.%

%TCIMACRO{\FRAME{ftbpFU}{4.293in}{5.604in}{0pt}{\Qcb{Fig.1 The qualitative
%picture of real space criteria of magnetic state stability in different
%systems. The local and itinerant moment regions are shown. N=0 corresponds to
%local moment criteria S$_{0}$ (2). S$_{\infty}$ in FM case corresponds to the
%regular Stoner criteria of ferromagnetism.}}{}{fig1.eps}%
%width 4.293in;  height 5.604in;  depth 0pt;  original-width 7.8339in;
%original-height 10.2424in;  cropleft "0";  croptop "1";  cropright "1";
%cropbottom "0";  filename 'Fig1.eps';file-properties "XNPEU";}} }%
%BeginExpansion
\begin{figure}
[ptb]
\begin{center}
\includegraphics[
height=5.604in,
width=4.293in
]%
{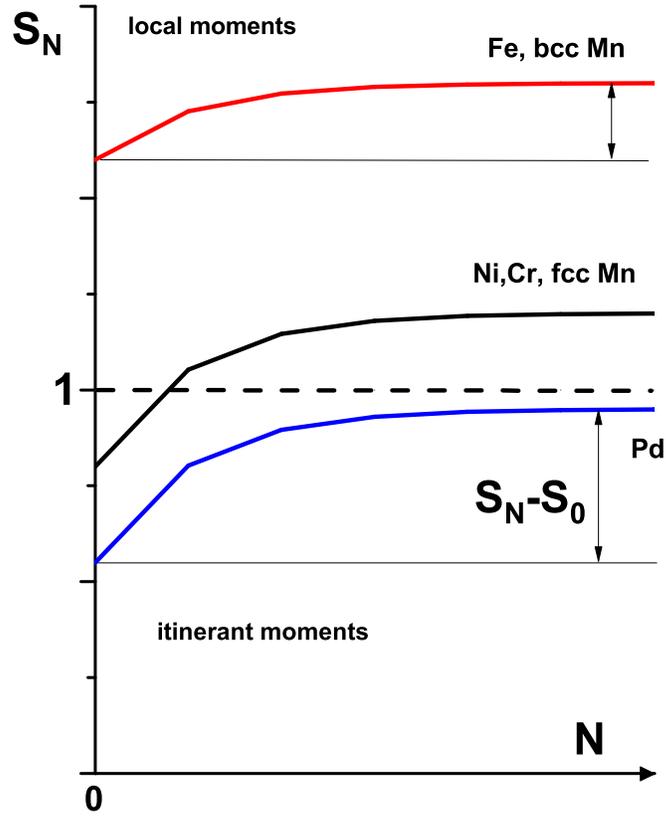}%
\caption{Fig.1 The qualitative picture of real space criteria of magnetic
state stability in different systems. The local and itinerant moment regions
are shown. N=0 corresponds to local moment criteria S$_{0}$ (2). S$_{\infty}$
in FM case corresponds to the regular Stoner criteria of ferromagnetism.}%
\end{center}
\end{figure}
%EndExpansion

%

%TCIMACRO{\FRAME{ftbpFU}{4.293in}{5.604in}{0pt}{\Qcb{Fig.2 The schematic view
%of the real space criteria of magnetic state stability for CaFe$_{2}$As$_{2}$.
%The lower curve corresponds to structure with small R$_{Fe\text{-}As}$ (in
%\AA ), the middle curve - "collapsed" tetragonal\cite{Caexp} (R$_{Fe\text{-}%
%As}$=2.336 \AA ) phase, and the upper curve corresponds to ambient pressure
%normal (R$_{Fe\text{-}As}$=2.373 \AA ) phase.}}{}{fig2.eps}%
%width 4.293in;  height 5.604in;  depth 0pt;  original-width 7.8339in;
%original-height 10.2424in;  cropleft "0";  croptop "1";  cropright "1";
%cropbottom "0";  filename 'Fig2.eps';file-properties "XNPEU";}} }%
%BeginExpansion
\begin{figure}
[ptb]
\begin{center}
\includegraphics[
height=5.604in,
width=4.293in
]%
{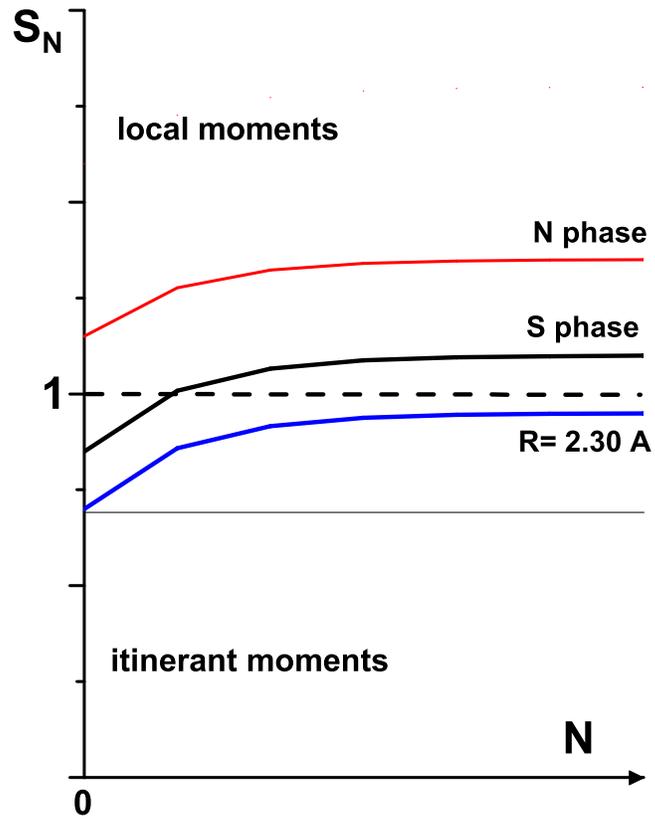}%
\caption{Fig.2 The schematic view of the real space criteria of magnetic state
stability for CaFe$_{2}$As$_{2}$. The lower curve corresponds to structure
with small R$_{Fe\text{-}As}$ (in \AA ), the middle curve - "collapsed"
tetragonal\cite{Caexp} (R$_{Fe\text{-}As}$=2.336 \AA ) phase, and the upper
curve corresponds to ambient pressure normal (R$_{Fe\text{-}As}$=2.373 \AA )
phase.}%
\end{center}
\end{figure}
%EndExpansion
%

%TCIMACRO{\FRAME{ftbpFU}{7.0668in}{6.5612in}{0pt}{\Qcb{Fig.3. The static
%susceptibility $\chi\left(  \QTR{bf}{q}\right)  $ of CaFe$_{2}$As$_{2}$ for
%the structures corresponding to the normal and superconducting
%phases\cite{Caexp}. a. $\chi\left(  \QTR{bf}{q}\right)  $ in plane [101] for
%the normal phase at ambient pressure. b. $\chi\left(  \QTR{bf}{q}\right)  $ in
%plane [110] for "collapsed" (superconducting) phase\cite{Caexp}. c.
%$\chi\left(  \QTR{bf}{q}\right)  $ in plane [110] for the normal phase at
%ambient pressure. The stability criteria corresponds to $1/I$=29.4 Ry$^{-1}$
%(see, Ref.\cite{Janak}). The vectors correspond to the structure rotated by 45
%degrees.}}{}{fig3a.ps}{\special{ language "Scientific Word";  type "GRAPHIC";
%width 7.0668in;  height 6.5612in;  depth 0pt;  original-width 6.7198in;
%original-height 6.2366in;  cropleft "0";  croptop "1";  cropright "1";
%cropbottom "0";  filename 'Fig3a.ps';file-properties "XNPEU";}} }%
%BeginExpansion
\begin{figure}
[ptb]
\begin{center}
\includegraphics[
height=6.5612in,
width=7.0668in
]%
{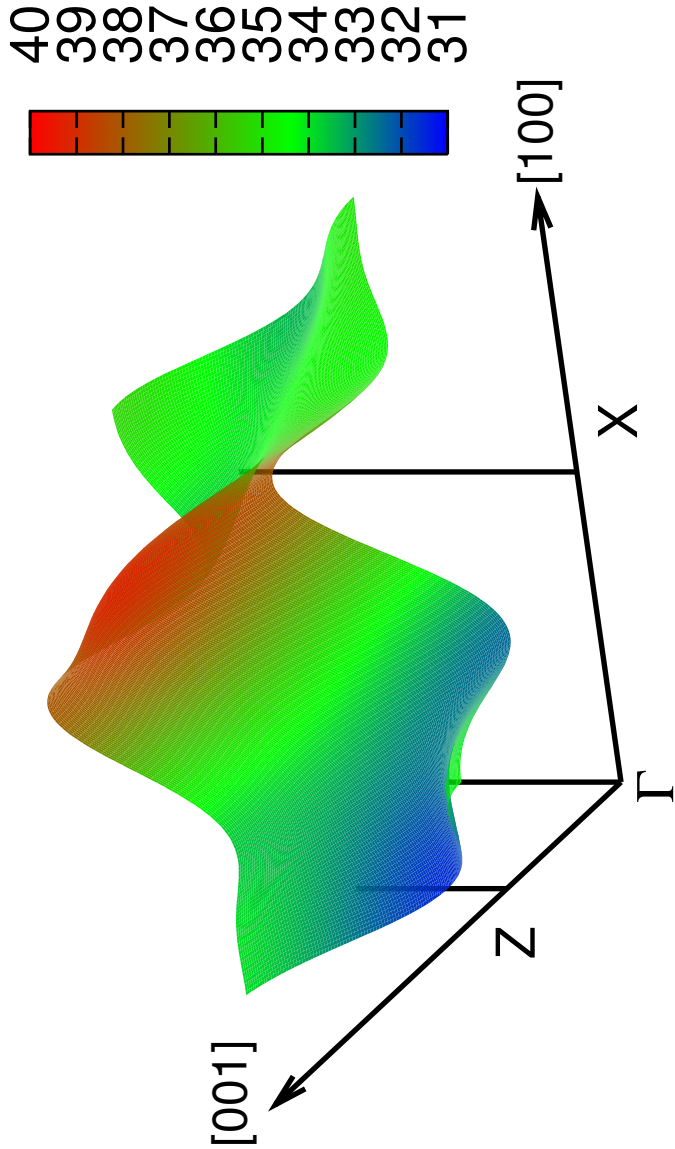}%
\caption{Fig.3. The static susceptibility $\chi\left(  \mathbf{q}\right)  $ of
CaFe$_{2}$As$_{2}$ for the structures corresponding to the normal and
superconducting phases\cite{Caexp}. a. $\chi\left(  \mathbf{q}\right)  $ in
plane [101] for the normal phase at ambient pressure. b. $\chi\left(
\mathbf{q}\right)  $ in plane [110] for "collapsed" (superconducting)
phase\cite{Caexp}. c. $\chi\left(  \mathbf{q}\right)  $ in plane [110] for the
normal phase at ambient pressure. The stability criteria corresponds to
$1/I$=29.4 Ry$^{-1}$ (see, Ref.\cite{Janak}). The vectors correspond to the
structure rotated by 45 degrees.}%
\end{center}
\end{figure}
%EndExpansion

%

%TCIMACRO{\FRAME{ftbpF}{7.0668in}{6.5604in}{0in}{}{}{fig3b.ps}%
%width 7.0668in;  height 6.5604in;  depth 0in;  original-width 6.7198in;
%original-height 6.2366in;  cropleft "0";  croptop "1";  cropright "1";
%cropbottom "0";  filename 'Fig3b.ps';file-properties "XNPEU";}} }%
%BeginExpansion
\begin{figure}
[ptb]
\begin{center}
\includegraphics[
height=6.5604in,
width=7.0668in
]%
{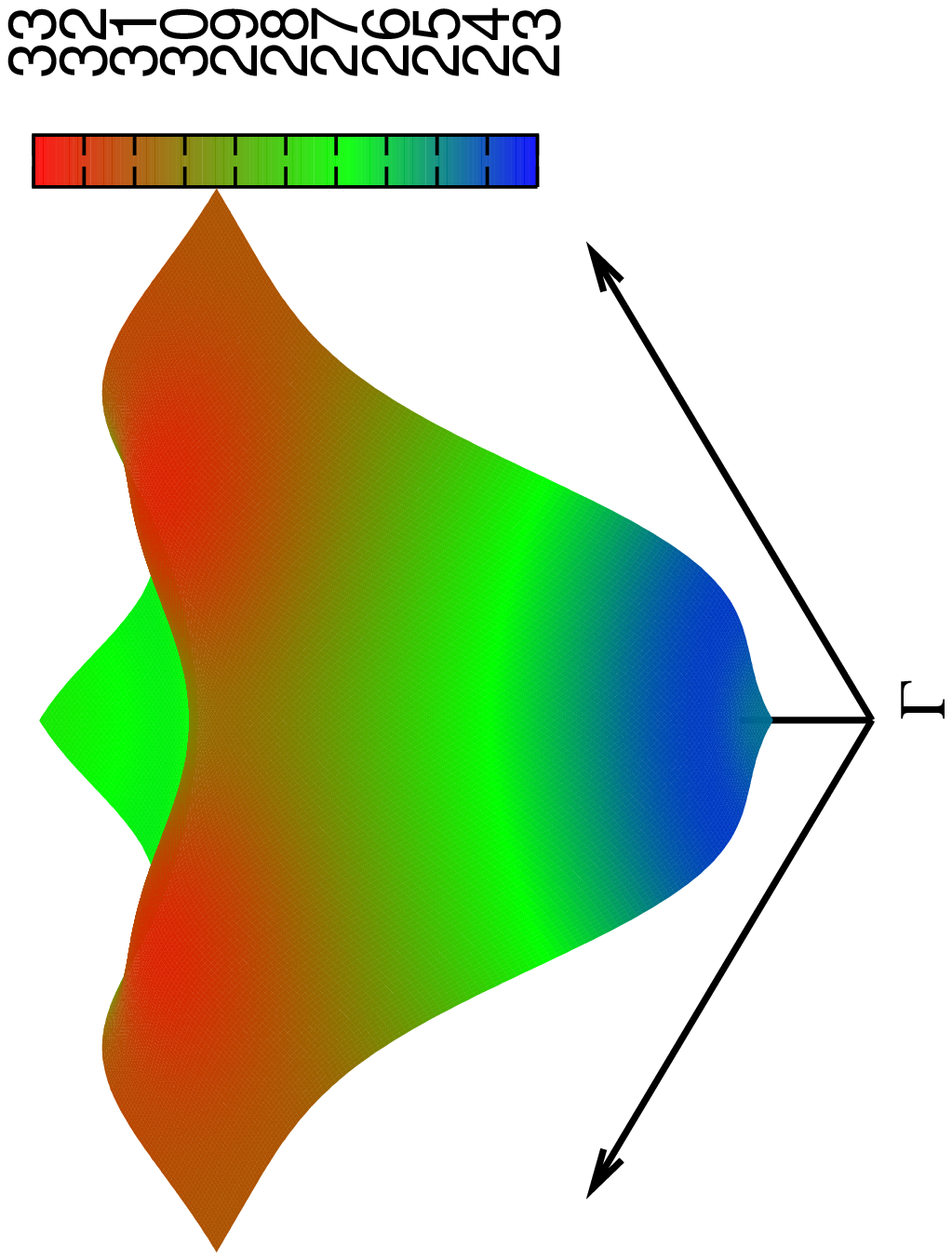}%
\end{center}
\end{figure}
%EndExpansion
%

%TCIMACRO{\FRAME{ftbpF}{7.0668in}{6.5604in}{0in}{}{}{fig3c.ps}%
%width 7.0668in;  height 6.5604in;  depth 0in;  original-width 6.7198in;
%original-height 6.2366in;  cropleft "0";  croptop "1";  cropright "1";
%cropbottom "0";  filename 'Fig3c.ps';file-properties "XNPEU";}} }%
%BeginExpansion
\begin{figure}
[ptb]
\begin{center}
\includegraphics[
height=6.5604in,
width=7.0668in
]%
{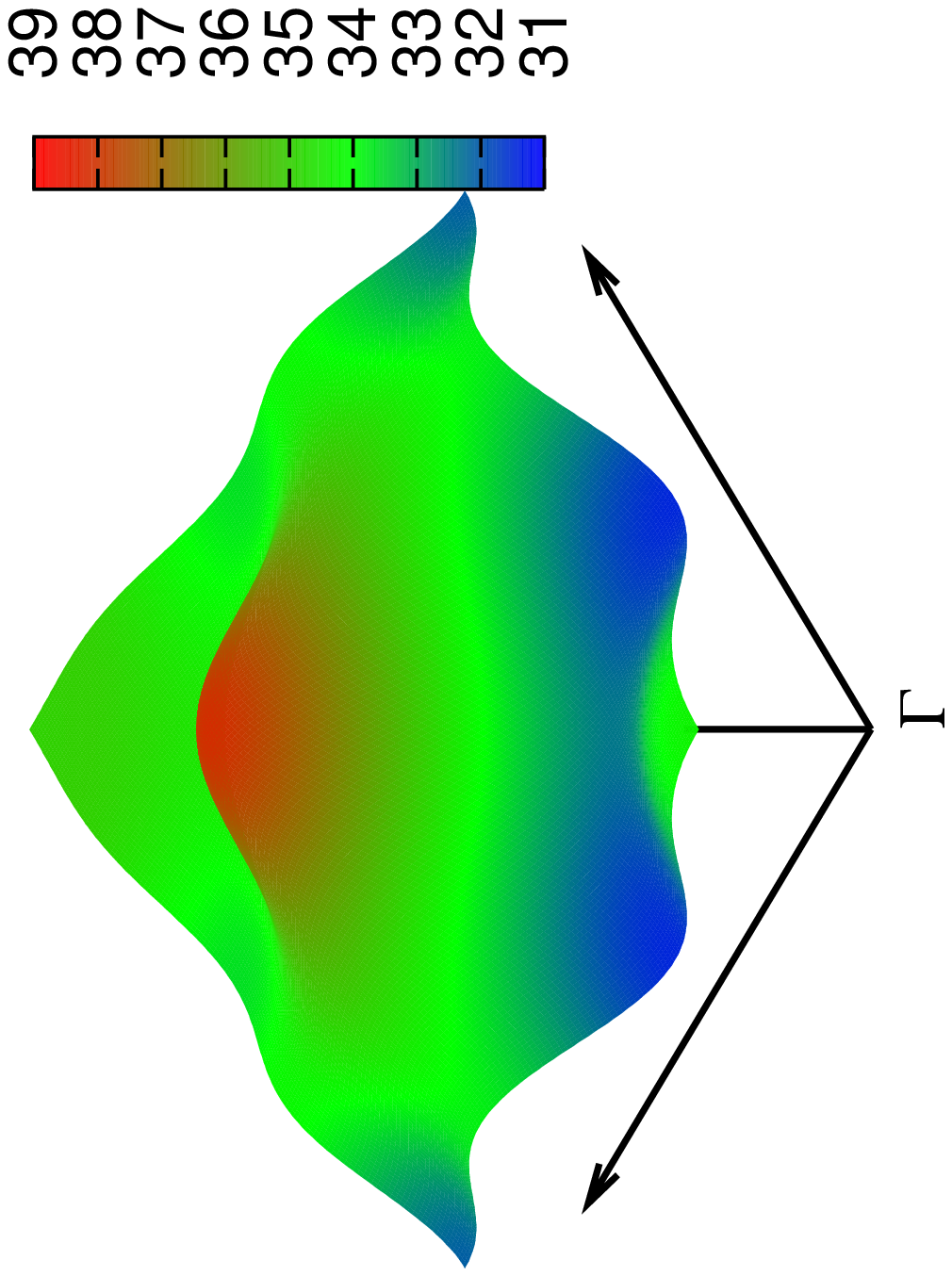}%
\end{center}
\end{figure}
%EndExpansion

\bigskip

\bigskip

\bigskip
\end{document}